\documentclass[aps,pre,twocolumn,showpacs]{revtex4-1}
\usepackage{graphicx}
\begin{document}
\title{Exact solution of rectangular Ising lattice in a uniform external field}
\author{C. B. Yang}
\affiliation{Key Laboratory of Quark \& Lepton Physics (MOE) and Institute of Particle Physics, Central China Normal
University, Wuhan 430079,  China\\
\small E-mail: cbyang@mail.ccnu.edu.cn
}
\date{\today}

\begin{abstract}
A method is proposed for exactly calculating the partition function of a rectangular Ising lattice with the
presence of a uniform external field. This approach is based on the method of the transfer matrix developed
about seventy years ago for the rectangular Ising model in the absence of external field. The basis for the vector
space is chosen as the eigenvectors of the diagonal part of the transfer matrix. The matrix elements for the
non-diagonal part can be calculated very easily. Then the partition function and thermodynamical quantities
can be evaluated. The limit of infinite lattice is discussed.

PACS number(s): 05.50.+q, 64.60.an, 75.30.Kz
\end{abstract}
\maketitle

The study of various phase transitions has been an extremely important research subject in many fields, because
phase transitions are common in physics and familiar in everyday life, as
liquid water freezes at zero census degree, the formation of binary alloys and the phenomenon of ferromagnetism.
In spite of their familiarity, phase transitions are not well understood.
The Ising model \cite{Ising}, which was initially proposed to explain how short-range interactions
give rise to long-range, correlative behavior, and to predict in
some sense the potential for a phase transition or spontaneous magnetization,
has been one of the most important models in investigating phenomena involving a phase transition. There is no other
local model that incorporates a phase transition that can be analyzed at anything like the resolution that is possible
for the Ising model \cite{Palmer}.
The Ising model has attracted enormous interest and also been applied to problems in chemistry, molecular biology,
and other areas where ``cooperative'' behavior of large systems is studied, and more than thousand
research papers published on properties of systems described by the model.

The one-dimensional Ising model model, which was suggested by Ernst Ising in the early 1920s, can be solved easily
but does not exhibit a  phase transition at any finite temperature. The interest in the model was revived when in 1936
R. Peierls argued \cite{peierls} that the two-dimensional Ising model should have a phase transition.
The transition point was located by Kramers and Wannier in 1941 \cite{kramers}.
The exact solution for the partition function of the rectangular Ising model was obtained in 1944 by
L. Onsager \cite{onsager} by using the transfer matrix method introduced in \cite{montr,kramers},
when there is no external field interacting on the lattice system. In 1949 B. Kaufman simplified Onsager's
calculations \cite{kaufman}. Since then, many different methods have been developed
for studying the thermodynamical properties of the Ising model in the absence of external field.
For a recent review,
 one can read \cite{Palmer,hystad}. Up to now, however, the two-dimensional Ising model in the presence
of a uniform external field has not been solved exactly, and Monte Carlo simulation is the only way to study the
properties of the Ising lattice under the influence of external field. Such simulation is powerful only for
lattices of small sizes because of the limitation of computer memory and takes a long time for computing
physical quantities to a high
accuracy. Because of the extremely wide applications of the model, finding an exact solution of the in-field  model
is very important and may bring us deeper understanding of the model and the phenomenon
of magnetization and order-disorder transitions.

In this paper, based on the results of \cite{onsager} and \cite{kaufman}, a method is proposed for calculating
exactly the partition function of the rectangular Ising model in the presence of a uniform external field.
The transfer matrix for an $N\times M$ lattice is represented by a  $2^N\times 2^N$ matrix. The partition
function of the system can be obtained by calculating the maximum eigenvalue of the  matrix.
No approximation is made in the process, thus the solution obtained is exact for $M$ larger enough.

For an $N\times M$ rectangular Ising lattice in a uniform external field, the Hamiltonian of the system
can be written as
\begin{equation}
\varepsilon=-\sum_{i,j}(J_1\sigma_{i,j}\sigma_{i+1,j}+J_2\sigma_{i,j}\sigma_{i,j+1})-H\sum_{i,j} \sigma_{i,j}\ ,
\end{equation}
with the coupling constants between spins on the nearest neighboring sites $J_1>0, J_2>0$ and a uniform
external magnetic field $H>0$. In the above equation, any $\sigma_{i,j}$ can take only two values, $\pm 1$.
The partition function can be expressed in terms of the transfer matrices as,
in a way similar  to that used in \cite{onsager},
\begin{equation}
Z=(\exp K_2)^{MN}{\rm Tr}(V_1V_2V_3)^M\ ,
\label{parti}
\end{equation}
where
\begin{eqnarray}
V_1&=&\prod_{j=1}^N (1+tC_j)\ ,\ \  t=\exp(-2K_2)\ ,\\
V_2&=&\prod_{j=1}^N \exp(K_1\sigma_j\sigma_{j+1})\ ,\\
V_3&=&\prod_{j=1}^N \exp(h \sigma_j)\ .\\
\end{eqnarray}
In the above expressions, $h= \beta H, K_i=\beta J_i$ for $i=1,2$.
Here $\beta=kT$ and $\sigma_j=\left(\begin{array}{cc}1 & 0\\ 0 & -1\\ \end{array}\right)$ and
$C_j=\left(\begin{array}{cc}0 & 1\\ 1 & 0\\ \end{array} \right)
$ are operators acting on the spin of site $j$ in a row.
When acting on a state function, the operator $\sigma_j$ gives us the spin ($\pm 1$) for the $j$-th site in a row,
but $C_j$ will reverse the spin of that site.
The operators $C_j$ and $\sigma_j$ satisfy the following quaternion algebra
 relations
\begin{eqnarray*}
\sigma_i\sigma_j&=&\sigma_j\sigma_i,\ \ C_iC_j=C_jC_i,\\
C_i\sigma_j&=&(-1)^{\delta_{i,j}} \sigma_jC_i\ , \ \ C_j^2=\sigma_j^2=1 \ .
\end{eqnarray*}
The partition function can be written in terms of the eigenvalues $\lambda_i$ of $V_1V_2V_3$ as $Z=(\exp(K_2))^{MN}
\sum_{i=1}^N \lambda_i^M$. When the lattice size $M$ is large enough, only the maximum eigenvalue $\lambda$ is
needed for  calculating the partition
function. In solving the eigenvalue problem, one can use a ``wrap-around'' model,
therefore, $\sigma_{N+1}=\sigma_1$.  In \cite{onsager,kaufman}, a chiral operator
$U=\prod_{j=1}^n C_j$ was defined and the matrix $V_1^{1/2}V_2V_1^{1/2}$ can be decomposed as a direct sum of
two parts corresponding to $U=1$ and $U=-1$, respectively.  Denote those eigenvectors of
$V_1^{1/2}V_2V_1^{1/2}$ corresponding to $U=1$ by $\Psi^+$, those corresponding to $U=-1$ by
$\Psi^-$. The former is even under operation of $U$, and the latter is odd.
Because $V_1^{1/2}V_2V_1^{1/2}$ is a Hermitian operator, the eigenvectors $\Psi^+$ and $\Psi^-$ form a complete
orthogonal basis for the vector space in question. Naively, a natural extension of the method to the case
with a uniform external field would be to calculate matrix elements of $V_3$ in the vector space spanned by $\Psi^+$
and $\Psi^-$. Because the operator $V_3$ does not commute with $U$, the acting of $V_3$ will cause a mixing
between states of $\Psi^+$ and $\Psi^-$. When the external field is weak, standard perturbation
theory can be used to obtain the partition function and the spontaneous magnetization \cite{cnyang}.
For this case, the only mixing needed for consideration is that between the two states $\Psi^+$ and $\Psi^-$
corresponding to maximum eigenvalues.  For the general case when the external field is not weak,
mixing among all states is possible and needed for solving the eigenvalue problem.
Then one has to work with a dense $2^N\times2^N$
matrix, thus analytical solution to the problem may be impossible for finite $M$ and $N$.
It will be shown in this paper that an exact solution can still be obtained.

To get the matrix elements for $V_1V_2V_3$, the eigenvectors $\Psi^+$ and $\Psi^-$ for $V_1^{1/2}V_2V_1^{1/2}$
are not a good choice as the basis, because those eigenvectors cannot be used easily in calculating the spin matrix
elements. If one wishes to
calculate the spin matrix of only one site, the site at the center of lattice for example, the method in
\cite{bugrij} can be used. We need, however, the spin matrix elements for all spins in a row, and
the method used in \cite{bugrij} does not work. Alternatively, to set up a basis,
one can first solve the eigenvalue problem for operator $V_2V_3$.
This is a simple problem, because that operator is diagonal in the meaning that it depends only
on the spin configuration of sites in a row. Thus this problem is like a one-dimensional Ising model.
One can identify an eigenvector of $V_2V_3$ by a set of sites with spin down, such as $|\emptyset\rangle$ for
the state with no spin down, $|2\rangle$ for the state with only one (the second site) spin
down, etc. If there are $n$ sites with spin down, there are $N!/n!(N-n)!$ possible ways to distribute those
sites in a row. Therefore, we have in total $2^N$ eigenvectors for the operator $V_2V_3$.
It is apparent that the set of such eigenvectors forms a complete orthogonal basis
for the problem involved. The eigenvalue of $V_2V_3$ corresponding to any one of those
states can be easily calculated, since it is determined only by the number of sites with spin
down ($N_-$) and the number of nearest neighboring down-down spin pairs ($N_{--}$). These two numbers can be
easily counted when the sites with spin down is fixed. Then the next task is to calculate
the matrix elements of $V_1$ on the basis. A crucial observation is that an eigenvector of $V_2V_3$ with $n$ sites
spin down, $|I_n\rangle$ with $I_n$ the set of sites with spin down, can be expressed as
\begin{equation}
  |I_n\rangle=\prod_{j\in I_n} C_j  |\emptyset\rangle\ .
  \end{equation}
  Because $C_i^2=1$ for any $i$, the action of $C_i$ on $|I_n\rangle$ may increase the number of spin down sites from
  $n$ to $n+1$ if the site $i$ is not included in $I_n$. Otherwise, $n$ will be decreased to $n-1$.
Therefore the operator $C_i$ can be an annihilation or a creation operator, depending on whether $i$ is
in the set $I_n$ or not. This observation makes the calculation of matrix elements for $V_1$ extremely easy.
To obtain the matrix elements for $V_1$, one first expands $V_1$ as
\begin{eqnarray}
V_1=\sum_{m=0}^N t^m\sum_{I_m} \prod_{j\in I_m} C_j\ ,
\end{eqnarray}
where $I_m$ is a set of $m$ different integers from 1 to $N$, and the summation over $I_m$ runs for all
possible different sets. Then a matrix element of $V_1$ between two states $|I_n\rangle$ and
$|I_l\rangle$, $\langle T_l|V_1|I_n\rangle$, is
\begin{equation}
V_1(n,l)=\sum_{m=0}^N t^m\sum_{I_m} \langle\emptyset|\prod_{i\in I_l}C_i\prod_{j\in I_m}
C_j\prod_{k\in I_n}C_k|\emptyset\rangle\ .
\end{equation}
Considering that the operator $C_j$ can be an annihilation or creation operator in different
situations, only one term in the above expression has nonzero value. That term has a special
property that in the set $I=I_l\bigcup I_m\bigcup I_n$, each site involved appears exactly twice. In other words,
the nonzero term in the above equation has $m$ equal to the number of different sites in sets $I_l$ and
$I_n$, because any site can be in $I_l$ and/or $I_n$ at most once. Therefore, $V_1(n,l)=t^m=V_1(l,n)$
with $m$ uniquely determined by the difference of sets $I_n$ and $I_l$, $m=n+l-2({\rm number\ of\ common\ sites})$.
So $V_1(n,n)=1$ for any $n$. Then the eigenvalue problem for the rectangular Ising model in a unifrom external field
\begin{equation}
(V_2V_3)^{1/2}V_1(V_2V_3)^{1/2}\Psi=\lambda \Psi\ ,
\end{equation}
can be rewritten as
\begin{equation}
BA=\lambda A\ ,
\label{neweig}
\end{equation}
where all the elements of $B$ are positive
\begin{equation}
B_{l,n}=B_{n,l}=\mu_l t^m \mu_n\ ,
\label{matr}
\end{equation}
with $\mu_l=\exp(K_1(N-4N_-+4*N_{--})/2+h(N-2N_-)/2)$ the eigenvalue of $(V_2V_3)^{1/2}$ corresponding
to $|I_l\rangle$. The vector $A$ in Eq. (\ref{neweig}) is for the expanding coefficients of $\Psi$
on the basis of $I_n$. There are effective ways for calculating the eigenvalue $\lambda$ with maximum magnitude
for such a symmetric matrix. From $\lambda$ all thermodynamical quantities can be calculated.

In this paper, we only consider the case with $J_1=J_2=J$, thus $K_1=K_2=K$.
We first investigate the temperature dependence of the mean spin per site $\langle\sigma\rangle=
\overline{\sum_{ij}\sigma_{ij}}/MN$ for an arbitrary chosen field $H=0.1J$.
It is obvious that
\begin{equation}
\langle\sigma\rangle=\frac{\partial \ln Z}{\partial h}/MN=\frac{\partial \ln \lambda}{\partial h}/N\ .
\end{equation}
Since the temperature $T$ appears in the problem always together with $J$ and $H$, one can
get the $T$ dependence of $\langle\sigma\rangle$ from its $K=\beta J$ dependence, which is shown
in Fig. 1. With the decrease of temperature or increase of $K$ from 0 to 1, $\langle\sigma\rangle$ increases
smoothly from 0 to 1, very quickly in the small $K$ region and saturating slowly in the low temperature (large $K$)
region. For comparison with the spontaneous magnetization for the field free situation \cite{mccoy},
the $K$ dependence of $\langle \sigma\rangle$ at $h=0$ for an infinite rectangular Ising lattice
is drawn also in Fig.1. At high temperature (or small $K$),
the external field makes $\langle\sigma\rangle$ larger than zero while the spontaneous magnetization is
nonzero only for $T<T_C$ or $K>K_C$. When the temperature is low enough,
the difference in $\langle\sigma\rangle$ for the two cases is very small, because almost all spins have been
aligned to the direction of the external field without external field. It is obvious in the figure
that the presence of external field makes the behavior
of $\langle\sigma\rangle$ analytic, very different from that at $H=0$ .
\begin{figure}[tbhp]
\centering
\includegraphics[width=0.45\textwidth]{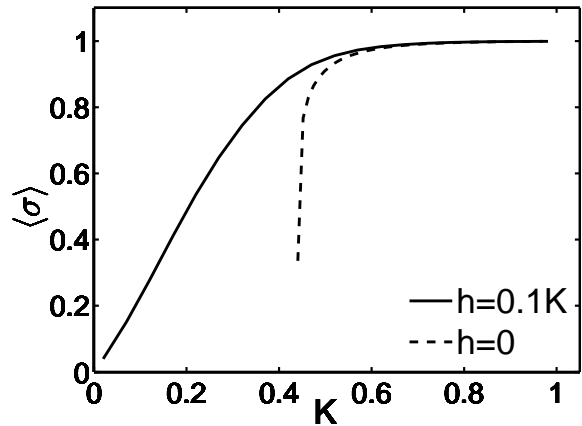}
\caption{ The mean magnetization per site $\langle\sigma\rangle$ as a function of parameter $K=\beta J$ for fixed
$H=0.1J$. The dashed curve for the case $h=0$ is drawn according to Eq. (4.9) on
page 245 of Ref. \cite{mccoy}.}
\end{figure}

We are more interested in the dependence of quantities on the external field $H$. Therefore, one can study
the $H$ dependences of $\langle\sigma\rangle$ and the mean energy per site. For this purpose, we fix the coupling
$J_1=J_2=k$ as an example, and investigate the dependence of $\langle\sigma\rangle$ on $H/k$ for a few
temperatures T=1.0, 2.0, 4.0 and 6.0. The results are shown in Fig. 2.
At low temperature $T=1.0$, which is well below the critical temperature
for the case with $H=0$, $\langle\sigma\rangle$ is almost 1 even for very weak external field. With the increase
of $T$, $\langle\sigma\rangle$ becomes smaller in low $H$ region and increases with $H$.
\begin{figure}[tbhp]
\centering
\includegraphics[width=0.45\textwidth]{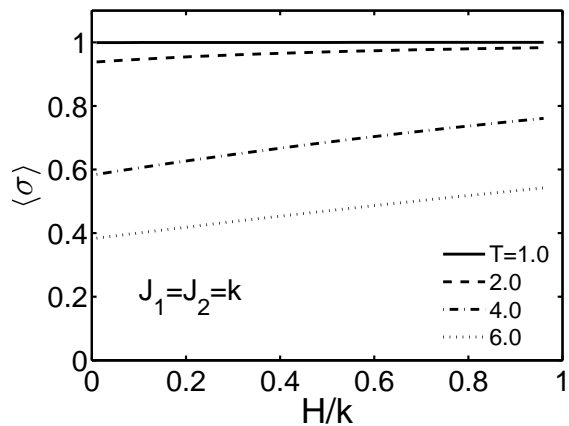}
\caption{ The mean magnetization per site $\langle\sigma\rangle$ as a function of  $H/k$ for fixed
$J_1=J_2=k$ at four temperatures.}
\end{figure}
One can get the mean energy $E$ per site from
\begin{equation}
-\beta E=\left(K\frac{\partial \ln Z}{\partial K}+h\frac{\partial \ln Z}{\partial h}\right)/(MN)\ .
\end{equation}
The numerical results for $-\beta E$ are shown in Fig. 3.
\begin{figure}[tbhp]
\centering
\includegraphics[width=0.45\textwidth]{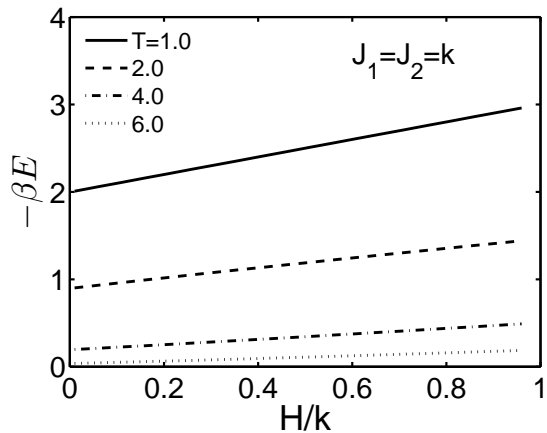}
\caption{ The product of $-\beta$ and the mean energy per site $E$ as a function of  $H/k$ for fixed coupling
$J_1=J_2=k$ at four temperatures as in Fig. 2.}
\end{figure}
At the four temperatures as in Fig.2, the product $-\beta E$ depends approximately linearly on the external
field $H$ in the region shown. This can be understood in combination with the $K$ dependence of $\langle\sigma
\rangle$ in Fig.1. Even at $T=4.0, K=\beta J=0.25$, $\langle\sigma\rangle$ is about 0.7. Thus the increase of
$-\beta E$ from the spin-spin interaction is very small with the increase of $H$. The increase of $-\beta E$  with
$H$ comes mainly from the field-spin interaction term which is proportional to the strength $H$ of the external
field. At low temperature, the mean energy is not zero at zero external field, because of the spontaneous
magnetization.  At $H=0$, the lower the temperature, the more the aligned spins, the lower the energy.

\begin{figure}[tbhp]
\centering
\includegraphics[width=0.45\textwidth]{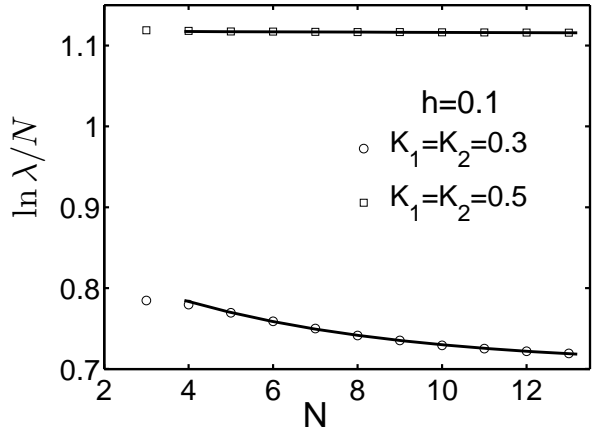}
\caption{ The lattice size $N$ dependence of the maximum eigenvalue $\lambda$ at two cases with
different values of $K=0.5$ and 0.3 but with fixed external field $h=0.1$. The solid curve is from fitting
the results to Eq. (\ref{fit}).}
\end{figure}

The method described in this paper can be used for finite $N$, whereas the value of $M$ can be arbitrarily large.
The matrix involved in calculating the maximum eigenvalue is $2^N\times 2^N$, thus its dimension increases very fast
with $N$. For real applications, one needs to study the thermodynamical limit, $N\to\infty$ and $M\to\infty$.
The $N$ dependence of the maximum eigenvalue $\lambda$ is shown in Fig. 4, for two cases, one with $K_1=K_2=0.5$, the
other with $K_1=K_2=0.3$, while the external field is fixed at $h=0.1$. For larger $K_1=K_2$, $\ln\lambda/N$ is
larger. For the case with higher $K_1=K_2$, the lattice size $N$ dependence of
$\ln\lambda/N$ is weaker. One can see that, with the increase of lattice size $N$, $\ln\lambda/N$ decreases and
approaches its saturation value quickly. In fact, points shown in Fig.4
for the two cases can be well described by
\begin{equation}
\ln\lambda/N=a+b\exp(-cN)\ ,
\label{fit}
\end{equation}
with the saturation value $a=1.106$ for the case with $K_1=K_2=0.5$ and 0.704 for the other case.
The fitted value of the parameter $c$ equals to 0.187 for the smaller $K_1=K_2$ case, while it is
$5.69\times 10^{-5}$ for the other case. Similarly, the lattice size dependence and
the infinite $N$ limit for thermodynamical quantities can be obtained.

The method developed in this paper can be extended to situations much more complicated.
When $J_2$ depends on the position of column, $J_2$ in Eq. (1) must be replaced by $J_{2j}$.
For this case, the only modifications are replacements of $\exp(NK_2)$ in Eq. (\ref{parti})
by $\prod_j \exp(K_{2j})$ and $t^m$ in Eq. (\ref{matr}) by $\prod_j t_j$. When the coupling $J_1$
 depends on the position $j$, the method in this paper can also be used with a modification $K_1\to K_{1j}$.
In this case, the eigenvalues $\mu_l$ and $\mu_n$ in Eq. (\ref{matr}) cannot be expressed simply in terms
of $N_-$ and $N_{--}$ only, but depend on the partition of the spin down-down pairs to a row.
If the external field is fixed but not uniform, similar extension can also be made.

In summary, we proposed a method for exactly calculating the partition function of the rectangular
Ising model with the presence of a uniform external field. With suitably chosen basis, the elements of the
transfer matrix and the maximum eigenvalue can be evaluated without any approximation.
The temperature and field strength dependence of the mean magnetization and mean energy per site are
presented. Though this method can be used only for a lattice with finite size in one direction,
the infinite limit can be obtained from the lattice size dependence of the thermodynamical quantities.
Applications of the method to much more complicated situations are straightforward.

This work was supported in part by the National Natural Science Foundation of China under Grant Nos.
11075061 and 11221504, by the Ministry of Education of China under Grant No. 306022 ,
and by the Programme of Introducing Talents of Discipline to Universities under Grant No. B08033.


\begin{thebibliography}{99}
\bibitem{Ising} E. Ising, Zeits. f. Physik, {\bf 31}, 253 (1925).
\bibitem{Palmer} J. Palmer, Prog. Math. Phys. {\bf 49} (Birkh\"auser Boston, 2007).
\bibitem{peierls} R. Peierlsm Proc. Camb. Phil. SOc. {\bf 32}, 477 (1936).
\bibitem{kramers} H.A. Kramers and G.H. Wannier, Phys. Rev. {\bf 60}, 263 (1941).
\bibitem{onsager} L. Onsager, Phys. Rev. {\bf 65}, 117 (1944).
\bibitem{montr} E. Montroll, J. Chen. Phys. {\bf 9}, 706 (1941).
\bibitem{kaufman} B. Kaufman, Phys. Rev. {\bf 76}, 1232 (1949).
\bibitem{hystad} G. Hystad, J. Math. Phys. {\bf 52}, 013302 (2011).
\bibitem{cnyang} C.N. Yang, Phys. Rev. {\bf 85}, 808 (1952).
\bibitem{bugrij} A.I. Bugrij and O. Lisovyy, Phys. Lett. {\bf A 319}, 390 (2003); J. Palmer and
G. Hystad, J. Math. Phys. {\bf 51}, 123301 (2010).
\bibitem{mccoy} B.M. Mccoy and T.T. Wu, The two-dimensional Ising model, {\it Harvard University Press, Canbridge,
Massachusetts}, 1973.
\end{thebibliography}
\end{document}